# 3D Vertical Dual-Layer Oxide Memristive Devices for Neuromorphic Computing

Siddharth Gaba, Patrick Sheridan, Chao Du, and Wei Lu*
Electrical Engineering and Computer Science, University of Michigan, Ann Arbor, MI

*Corresponding author. Electronic mail: wluee@eecs.umich.edu

*Abstract*— **Dual-layer resistive switching devices with horizontal W electrodes, vertical Pd electrodes and $WO_x$ switching layer formed at the sidewall of the horizontal electrodes have been fabricated and characterized. The devices exhibit well-characterized analog switching characteristics and small mismatch in electrical characteristics for devices formed at the two layers. The three-dimensional (3D) vertical device structure allows higher storage density and larger connectivity for neuromorphic computing applications. We show the vertical devices exhibit potentiation and depression characteristics similar to planar devices, and can be programmed independently with no crosstalk between the layers.**

*Index Terms*—3D memory, memristor, resistive switching, RRAM, neuromorphic computing.

.

## I. INTRODUCTION

To maintain functional scaling of integrated circuits, vertical scaling , that aims at enhancing the device performance or functionality through expansion in the vertical direction, is now being widely researched for future memory and logic applications[1]. In particular, resistive memories (RRAMs) based on two-terminal resistive switching devices have attracted broad attention [2]–[4] due to their compatibility with vertical scaling. Generally, two different approaches are being investigated for vertical scaling in resistive memory devices: traditional 3D cross-point RRAM [5] using stacked cross-point arrays in a layer-by-layer fashion; and vertical RRAM structures [6], [7] based on devices formed at the sidewall between a vertical electrode and a lateral electrode with the capability to form multiple layers simultaneously (Fig. 1a).

The vertical 3D RRAM structure or sidewall type of structure has several advantages over the traditional crosspoint-type of device structure. In a conventional cross point device, the active area dimensions are completely defined by lithography. As devices scale, the cost of lithography steps increases drastically. However, in sidewall devices at least one active device dimension is not critically dependent on lithography. The top (vertical) electrode is still defined by lithography in both conventional crosspoint devices and in sidewall devices. In the latter case, however, the active area dimension is determined by the thickness of a deposited film which can be precisely controlled to the atomic level, as opposed to the lithographically defined dimension of the bottom electrode. Additionally, since deposition thicknesses can be controlled to a much better extent than defined by lithography, device to device variation can be improved. Finally, for vertical RRAM structures the number of critical lithography steps remains fairly constant as the number of layers increases, implying significant improvements in cost savings compared with stacked cross-point approaches.

In this brief, we demonstrate that vertical, 3D dual layer $WO_X$-based resistive switching devices exhibit well-defined analog memristive switching behavior. The device characteristics in the dual layer stack are closely matched and exhibit excellent incremental potentiation and depression characteristics with no degradation up to ten thousand cycles. The demonstration of vertical, multi-layer memristive devices fabricated in CMOS friendly fashion makes them well suited for analog memory or large-scale neuromorphic computing applications.

## II. Device fabrication

The dual layer vertical devices were fabricated on a Si/SiO$_2$ substrate with 100nm of thermal oxide. The first horizontal electrode layer of tungsten (40nm) was deposited at room temperature by DC sputtering in a Kurt J Lesker LAB 18 system. Although many different metal oxides have been studied as candidates for memristive devices [8], tungsten-based materials were chosen for this demonstration due to the ubiquitous use of W in commercial CMOS processes and the rich knowledge of this material. Silicon dioxide (60nm) serving as the inter layer dielectric, was then deposited at 200°C in a plasma enhanced chemical vapor deposition system (GSI PECVD). Ellipsometric and X-SEM methods were utilized to control all thicknesses to within +/- 10% of nominal values. The tungsten and silicon dioxide depositions were then repeated to form the dual-layer horizontal electrode stack as shown in the Fig. 1a (inset). Next, photolithography and reactive ion etching (RIE) were used to pattern the film stack. To form the tungsten oxide (WO$_X$) switching layer, the sample was annealed in an oxygen rich ambient at 375°C at atmospheric pressure for 60 seconds in a JetFirst 150 RTP system. The exposed sidewalls were oxidized to form WO$_X$ while the remaining bulk of the tungsten, which was covered by the PECVD silicon dioxide, served as the horizontal electrodes (Fig. 1b). Afterwards, the vertical Pd electrodes were patterned through photolithography, e-beam evaporation and liftoff techniques to complete the Pd/WO$_X$/W device structure at each sidewall junction (Fig. 1b). To improve sidewall coverage of the top electrode, the sample was placed at an angle of ~45 degrees to the normal incident direction during electron beam evaporation of the vertical electrode material (Pd 600 Å / Au 2400 Å). Finally, photolithography and RIE were used to open contact pads to the two horizontal tungsten electrode layers, followed by Au pad deposition. Throughout the process the peak temperature was limited to at 375°C to maintain CMOS compatibility. A cross-sectional SEM image of a completed device is shown in Fig. 1c.

III. RESULTS AND DISCUSSIONS

The devices were tested in both DC and pulse operation modes using a custom-built test circuitry (Fig. 1d). Fig. 2 shows the I-V characteristics obtained from both the upper device and the lower device. A typical resistive switching behavior can be observed with well-defined hysteresis. This behavior is similar to results obtained from horizontal 2D devices [9], and verified the feasibility of the vertical device concept and proves that high quality $WO_x$ materials can still be obtained at the electrode sidewall reliably. The hysteresis in the I-V curves is attributed to the migration of oxygen vacancies in the non-stoichiometric $WO_x$ matrix [9]. During oxidation process, there are more oxygen vacancies generated near the outer surface (i.e. near the vertical Pd electrode side (Fig. 1d)). In this configuration, the total device resistance is dominated by the oxygen-vacancy poor region near the horizontal electrode. Applying a negative voltage to the W electrode drives the migration of the positively charged oxygen vacancies towards the W electrode and improves the overall device conductance. Conversely, applying a positive voltage to the W electrode drives the oxygen vacancies away towards the Pd electrode and thus makes the device less conductive. While horizontal devices rely on oxidation of a pristine as-deposited W interface to generate the $WO_x$ matrix with an oxygen vacancy concentration gradient, these vertical devices are obtained by oxidation of an etched sidewall. Thus, this vertical device concept proves that high quality $WO_x$ materials can still be obtained at the electrode sidewall reliably.

Significantly, very similar I-V curves can be obtained from both devices in different stacks along the same vertical electrode (Fig. 2a) indicating close matching between the two devices in the dual layer approach. The close matching is further demonstrated by comparing the current though each device while applying five consecutive negative DC set cycles (0V to -3V) followed by five consecutive positive DC reset cycles (0V to +3V) to each device (Fig. 2b). Each device demonstrates an incremental change in conductance as expected from an analog memristive device – gradual increase on applying a negative voltage and gradual decrease on applying a positive voltage, and very similar behavior can be observed in both devices (Fig. 2c). Fig. 2d plots the device current measured at the maximum programming voltage of

-3V during the 5 consecutive set cycles, demonstrating <10% mismatch between the upper layer and lower layer devices.

Analog memristive devices have been widely proposed to emulate synaptic functions in neuromorphic circuits [10]–[14]. In particular, large connectivity can be obtained when these devices are organized in a crossbar structure [15] that mimics the network structure of biological systems [10]. The ability to extend the network to the vertical direction in the 3D vertical structure demonstrated here will further improve the connectivity and allow large scale network developments. To this end, Fig. 3 shows the results from the top layer and bottom layer devices that demonstrate potentiation and depression synaptic behaviors when programmed with negative and positive pulse trains. Again, both devices in the dual layer structure show well-defined analog memristive behaviors and are closely matched (Fig. 3a, b). The potentiation and depression behaviors also remain stable as the devices are pulsed to 10000 cycles, with no degradation during the endurance test (Fig. 3c-f).

Finally, we show that each device in the dual layer stack can be programmed and read independently without any crosstalk with the non-addressed device, even though the devices are directly on top of each other and share the same vertical electrode. The two devices operate in parallel and thus can be conceptually seen as two independent devices sharing common bottom electrode, but with separate top electrodes. Alternatively, the vertically stacked memristive device array can effectively act as a single synapse to mitigate the limited resolution and the stochastic switching characteristics seen in single memristive devices.

Results from four different programming scenarios are shown in Fig. 4 showing independent control of the devices. For example, in the fourth scenario, the upper device was programmed with a -3V/400us pulse train (same as the condition used in Fig. 3). Switch S1 was then opened and the lower device was programmed by closing switch S2. The current level attained by the lower device in this case was then compared to that in scenario 2, where the upper device has not been programmed, and almost identical results were obtained demonstrating independent programming of each device.

IV. CONCLUSION

In summary, CMOS compatible, dual-layer vertical tungsten oxide resistive switching devices were demonstrated. The devices show well-defined incremental resistance switching behavior and good endurance exceeding 10,000 potentiation/depression cycles. The devices can be programmed with less than 10% mismatch and no apparent crosstalk. This scalable architecture is well suited for development of analog memory and neuromorphic systems. The conductance change ratio may be further increased by optimizing the stack etch, post etch cleans and the oxidation conditions and is the subject of further studies.

ACKNOWLEDGMENT

The authors thank Lin Chen, Dr. Yuchao Yang and Dr. Ting Chang for useful discussions. This work was supported in part by the Air Force Office of Scientific Research (AFOSR) through MURI grant FA9550-12-1-0038 and by the National Science Foundation (NSF) through grant CCF-1217972. This work used the Lurie Nanofabrication Facility at the University of Michigan, a member of the National Nanotechnology Infrastructure Network (NNIN) funded by NSF.

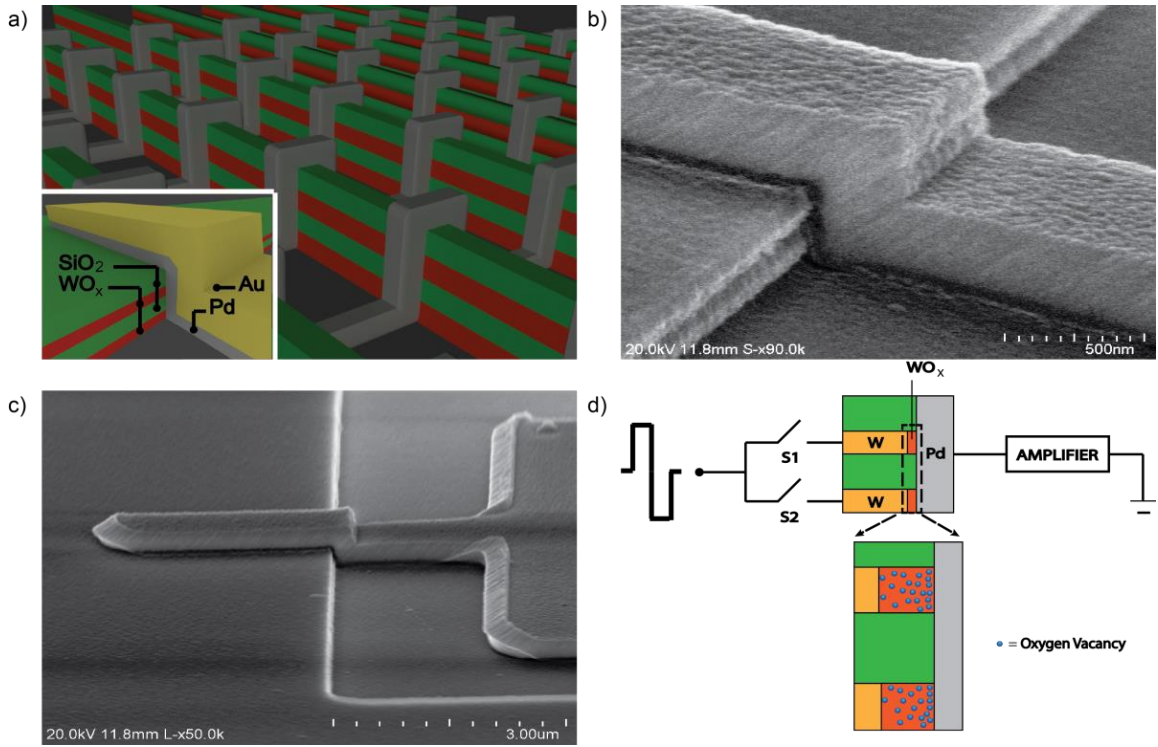

Fig. 1. (a) Schematic of vertical dual-layer RRAM array. Inset: Schematic of the fabricated devices. The WO$_X$ switching layers are formed at the sidewalls of the W electrodes (b) Scanning electron micrograph (SEM) image of the device. (c) SEM image of the complete device showing good sidewall contact. (d) Schematic of the test circuitry (top) and oxygen vacancy distribution in the switching layer (bottom).

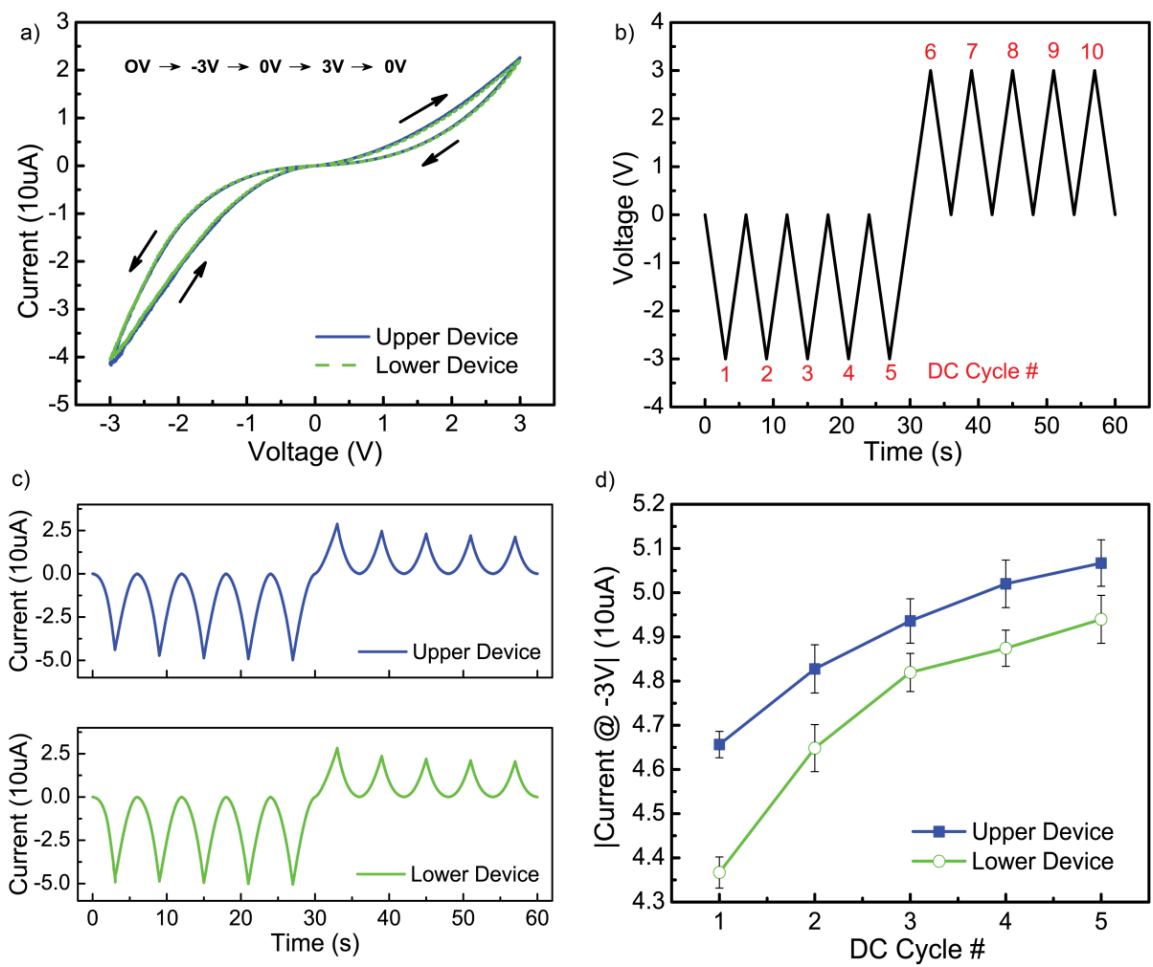

Fig. 2. (a) I-V plot measured for each device in the dual layer structure. (b) Input voltage waveform of five set cycles of 0 to -3V followed by five reset cycles of 0 to +3V at 1V/s. (c) Current output for the upper device(upper panel) and the lower device ( lower panel). (d) Incremental change of the maximum current during DC programming for both devices. The error bars were obtained from five different DC sweep measurements.

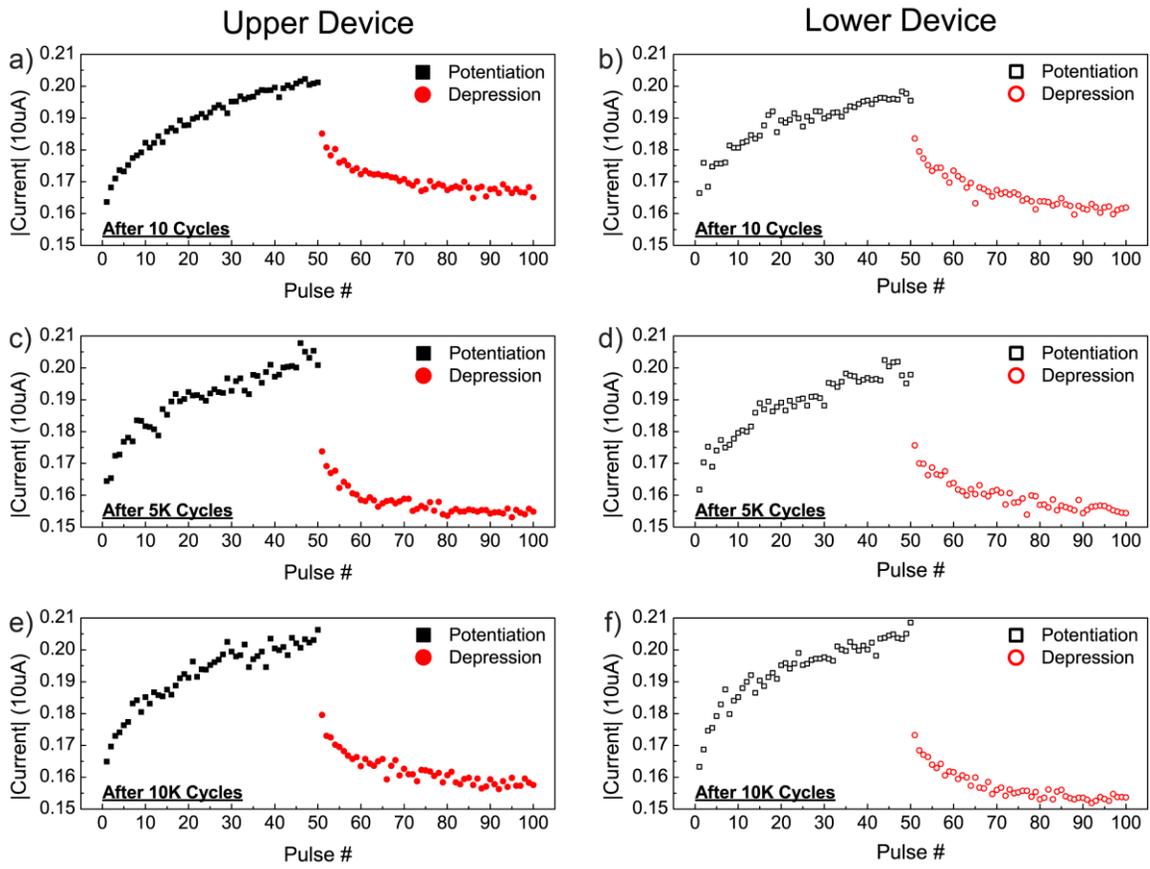

Fig. 3. Current measured after consecutive potentiation (-3V/400us) and depression pulses (3V/400us) for the upper device (a) and lower device (b). Each cycle comprises of 50 potentiation pulses and 50 depression pulses. The read voltage was -0.8V. The device performance remains unchanged after 5000 cycles (c, d) and 10000 cycles (e, f).

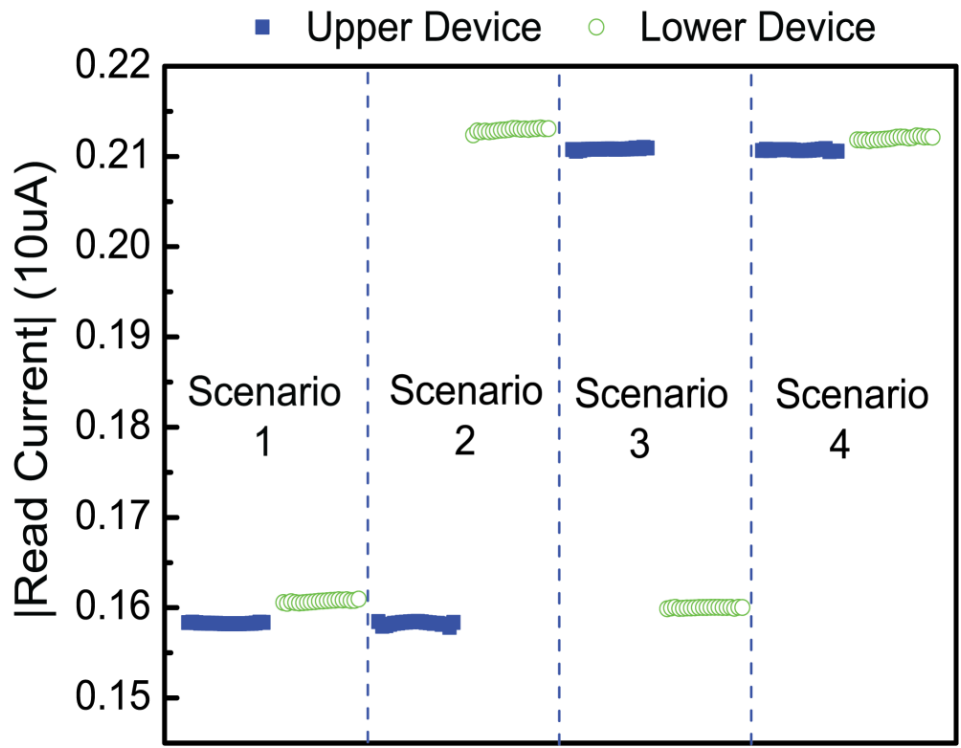

Fig. 4. Independent programing/read of devices in the dual-layer structure. The read currents obtained during 20 consecutive read pulses were plotted for the four scenarios after the upper/lower device has been reset/reset, reset/set, set/reset, and set/set, respectively. The devices were either set with 50 consecutive -3V/400us pulses or reset with 50 consecutive 3V/400us pulses. The read pulses were -0.8V/3ms.